\documentclass[aps,pre,twocolumn,superscriptaddress,floatfix,tightenlines,showpacs,notitlepage]{revtex4-1}
\usepackage{graphicx}
\usepackage{bm}
\usepackage[utf8]{inputenc}
\usepackage[english]{babel}
\usepackage{amsfonts}
\usepackage{color}
\usepackage{amsmath,amssymb}    
\usepackage{epsfig}
\usepackage{subfigure}  
\usepackage{hyperref}   
\usepackage{amssymb}
\usepackage{soul}
\usepackage[section]{placeins}
\usepackage{url}
\usepackage{ulem}
\AtBeginDocument{%
    \newwrite\bibnotes
    \def\bibnotesext{Notes.bib}
    \immediate\openout\bibnotes=\jobname\bibnotesext
    \immediate\write\bibnotes{@CONTROL{REVTEX41Control}}
    \immediate\write\bibnotes{@CONTROL{%
    apsrev41Control,author="08",editor="1",pages="1",title="0",year="1"}}
     \if@filesw
     \immediate\write\@auxout{\string\citation{apsrev41Control}}%
    \fi
}%

\newcommand{\ictsaddress}{International Centre for
  Theoretical Sciences, Tata Institute of Fundamental Research,
  Bangalore 560089, India}

\newcommand{\minesaddress}{MINES ParisTech, PSL Research University, CNRS, CEMEF, Sophia-Antipolis, France}
\newcommand{\inria}{Universit\'e C\^ote d'Azur, Inria, CNRS, Cemef, Sophia-Antipolis, France.}
\newcommand{\imscaddress}{The Institute of Mathematical Sciences, CIT Campus, Taramani, Chennai 600 113, India}
\begin{document}
\title{Turbulent Route to Two-Dimensional Soft Crystals}
\author{Mohit Gupta}
\email{mohit.gupta9607@gmail.com}
\affiliation{\ictsaddress}
\affiliation{School of Physics and Astronomy, University of Minnesota, Minneapolis, Minnesota 55455, USA}
\author{Pinaki Chaudhuri}
\email{pinakic@imsc.res.in}
\affiliation{\imscaddress}
\author{J\'er\'emie Bec}
\altaffiliation{Also: Associate, International Centre for Theoretical Sciences, Tata Institute of Fundamental Research, Bangalore 560089, India}
\email{jeremie.bec@mines-paristech.fr}
\affiliation{\inria}
\affiliation{\minesaddress}
\author{Samriddhi Sankar Ray}
\email{samriddhisankarray@gmail.com}
\affiliation{\ictsaddress}

\begin{abstract}
We investigate the effect of a two-dimensional, incompressible, turbulent flow
on soft granular particles and show the emergence of a
crystalline phase due to the interplay of Stokesian drag and short-range
inter-particle interactions. We quantify this phase through the bond order
parameter and local density fluctuations and find a sharp transition between
the crystalline and non-crystalline phase as a function of the Stokes number.
Furthermore, the nature of preferential concentration, characterised by the
correlation dimension, is significantly different from that of
particle-laden flows in the absence of repulsive potentials.
\end{abstract}

\maketitle

The self-assembly of particles in a flow \cite{whitesides,
  liz-marzan}, because of its ubiquity, is amongst the most studied
problem in the areas of turbulent transport, soft matter, granular
systems and nonequilibrium statistical mechanics.  In quiescent form,
most dilute assemblies are liquids, which when densified, can take a
crystalline or amorphous structure depending upon the dispersity of
the constituents \cite{Clark, pusey, liunagel, boyer}. Consequently,  
extensive studies of the rheology of such suspensions have
happened \cite{coussot,bonn}, motivated by diverse applications. In
typical experiments and computer simulations, the role of a carrier
flow in dispersing the particulate matter is trivial: indeed if there
is an underlying fluid medium, they are typically simple
shearing~\cite{qflow1,qflow2}.  In a variety of natural and industrial
processes, however, particles are dispersed in flows with non-trivial
spatio-temporal correlations which are chaotic, and in extreme cases
(such as a marine system) even turbulent~\cite{Wells1,Wells2,Kepskey}.

This specific question of the structural properties of particulate suspension,
where the underlying flow is turbulent, has surprisingly been not investigated
despite significant progress in the last two decades in the area of turbulent
transport of finite-sized, heavy, inertial (colloidal) particles.  In this
paper, we report the emergence of macroscopic particulate structures with
crystalline (hexagonal) motifs even in the presence of strong mixing because
of the carrier turbulent flow.  

\begin{figure}
  \begin{center}
    \includegraphics[width=1.\columnwidth]{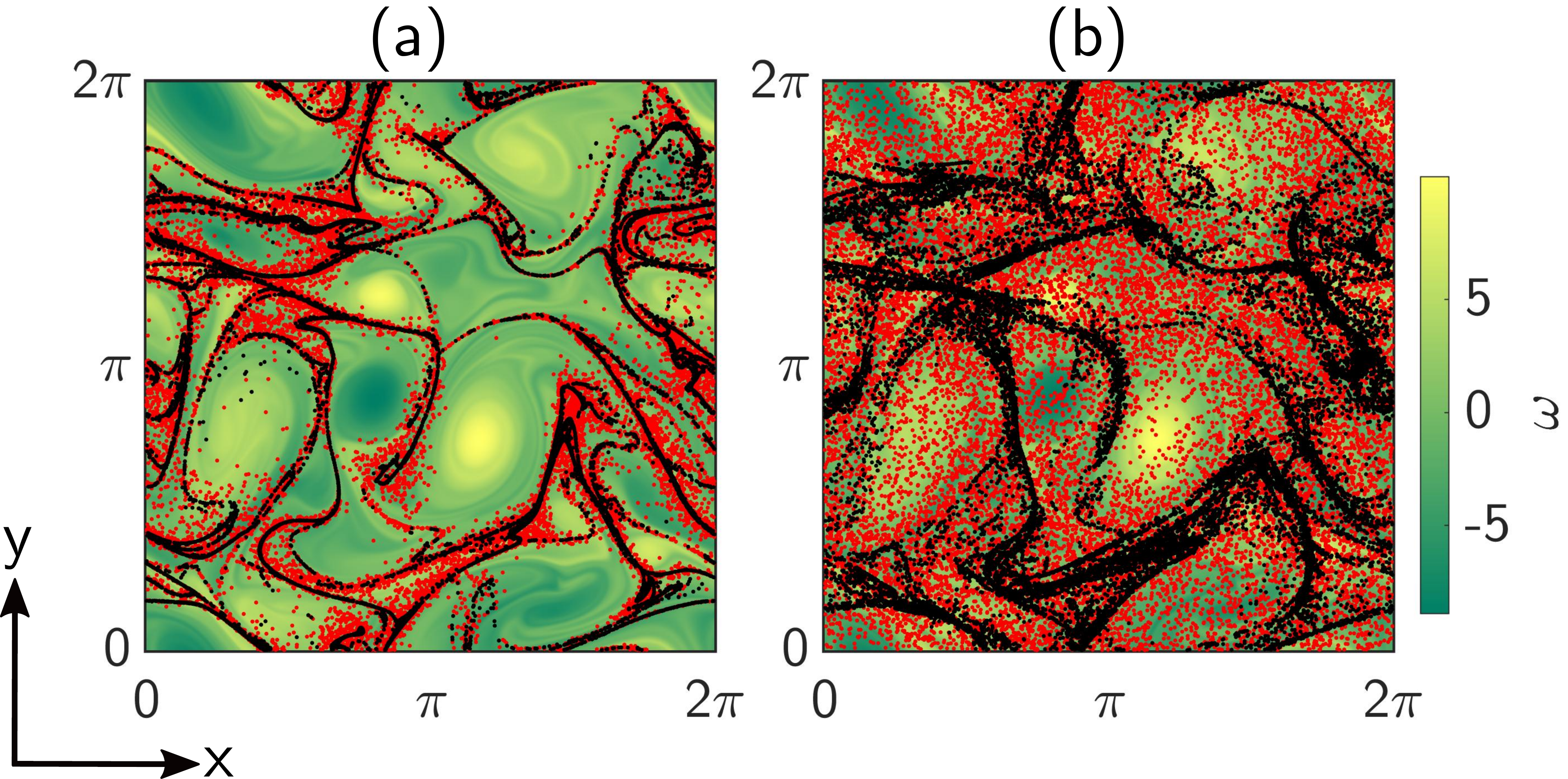}
    \caption{$\phi=0.1\%$. Representative snapshots of  non-interacting particles, i.e. $V(r_{ij})=0$ (shown in black) and
interacting soft particles (shown in red) with $V(r_{ij})$ specified in Eqn.\ref{potential}, for Stokes numbers (a) $St=0.75$ and 
      (b) $St=2.50$, superimposed on vorticity field of the carrier turbulent flow.}
    \label{traj}
  \end{center}
\end{figure}

For suspended particles with a finite diameter and density, inertial
effects and dissipative dynamics become important, leading to the
particles detaching from the underlying flow to form strong
inhomogeneities in their spatial distribution (see
Fig.~\ref{traj}). This phenomenon, known as
$\textit{preferential concentration}$, has been extensively
studied~\cite{pref-conc,monchaux,mehlig,inertial,goto-sweep-stick,collins-prefconc}
and remains critical to our explanations of problems such as rain
initiation in warm clouds.  Although more recent studies have
addressed the issue of effects such as
gravity~\cite{ssr-gravity,collins-gravity,gustavsson-gravity} and
turbophoresis \cite{turbo-1,turbo-2,turbo-3} on preferential
concentration, inter-particle interactions have largely been ignored
except for studies on coalescences~\cite{Bec-coal,Li-coal} and on
Vicsek ordering~\cite{RayEPL}. The only example that we are aware of,
which links ideas of soft matter and turbulent transport is the use of
repulsive, elastic, hard sphere inter-particle interactions, which,
combined with a dissipative dynamics, lead to \textit{stickiness} and
aggregation~\cite{sticky}.  However, for most physical systems, the
elastic limit is an idealised one: The most obvious particulate
exchanges, such as those mediated through a \textit{soft
  potential}~\cite{gran} has been ignored so far.

In this work, we therefore address two important and related issues, namely
what is the effect of soft particle interactions on clustering of particles in a
turbulent flow and can such realistic interactions, contrary to na\"ive
expectations, lead to the growth of stable crystalline structures in an
ensemble of particles interacting with each other as well as an ambient
turbulent fluid. 

We consider an assembly of $N_p$ particles seeded in a
two-dimensional, statistically stationary, turbulent velocity field
$\textbf{u}$~\cite{SI}.  Since we consider particles smaller than the relevant
length scales of the flow, the dynamics of the $i$-th particle
(characterised by its Stokes or particle response time $\tau_p = \frac{2\rho_p a^2}{9\rho_f\nu}$, where 
$a$ is the particle radius, $\nu$ the kinematic viscosity of the flow, and $\rho_p$ and $\rho_f$ are the particle and fluid 
densities, respectively)
defined through it position $\textbf{x}_i$ and velocity $\textbf{v}_i$
is given by the linear Stokes drag model along with the inter-particle
interaction potential $V(r_{ij})$:
\begin{equation}
  \frac{d\textbf{x}_i}{dt} = \textbf{v}_i;\quad
  \frac{d\textbf{v}_i}{dt} = -\frac{\textbf{v}_i -
    \textbf{u}(\textbf{x}_i,t)}{\tau_p} - \sum_{\substack{j=1\\ j\neq
      i}}^{N_p}{\bf \nabla}V(r_{ij});
  \label{stokes}
\end{equation}
where the interacting short-ranged repulsive potential
\cite{durian95,gran}, commonly used for modelling emulsions and other
soft granular suspensions, is given by:
\begin{equation}
  V_{\rm soft}(r_{ij})=
  \begin{cases}
    \frac{\epsilon}{2}(1-r_{ij}/\sigma_{ij})^2 & \text{for } r_{ij}< \sigma_{ij},\\
    0 & \text{for } r_{ij}\geq \sigma_{ij};
  \end{cases}
  \label{potential}
\end{equation}
where, $r_{ij}$ is the inter-particle separation, $\sigma_{ij}$ is the sum of the radii of particles $i$ and $j$,
$\epsilon = 1$ sets the energy scale for particle interactions. Thus, such an interaction takes into account
the energy cost of deformation, only when two particles are in contact. 
For our work, we mostly consider a mono-disperse assembly,  
where all particles have the same diameter $\sigma$. For the mono disperse suspensions, which is our focus, all particles 
have the same radius $a$; thence $\sigma=2a$. In the Supplemental Material~\cite{SI} we show additional results for bi-disperse 
suspension where particles have different radii.

The advecting turbulent velocity $\textbf{u}$ is obtained (by using a
standard pseudo-spectral method) as a solution of the two-dimensional,
incompressible Navier-Stokes equation, written in the 
vorticity ($\omega = \nabla \times {\bf u}$) formulation as
\begin{equation}
	\partial_t \omega + {\bf u}\cdot\nabla \omega = \nu\nabla^2\omega - \alpha \omega + {\bf f};
\end{equation}
with $\nu$ the kinematic 
viscosity and $\alpha$ Ekman friction coefficient.
The flow is solved on a 2$\pi$ periodic grid with $N=512^2$ collocation points, and
a deterministic large scale forcing $f$, ensuring a 
direct enstrophy cascade regime, maintains a non-equilibrium steady
state~\cite{RayPRL}. 
The flow is conveniently described by its characteristic length
$l_\nu=\sqrt{\langle\Omega\rangle/\langle P\rangle} = 0.06$ and time
$\tau_{f}=\sqrt{1/\langle\Omega\rangle} = 0.4$ scales, where
$\Omega = \int k^2 E(k) dk$ is the enstrophy and $P=\int k^4 E(k) dk$ the
palinstrophy~\cite{Boffeta}. We set the forcing length scale $l_f=1.71$ yielding a secondary time scale 
$l_f/u_{\rm rms}= 1.60$, $u_{\rm rms}=\sqrt{\int E(k)dk}$.
By using the flow time-scale, we define the
non-dimensional Stokes number $St = \tau_p/\tau_f$ and ensure that
both the grid spacing and particle diameter are much smaller than
$l_\nu$. Further, we vary the Stokes number by using different
values of $\tau_p$. 

\begin{figure}
  \centering
  \includegraphics[width=1\columnwidth]{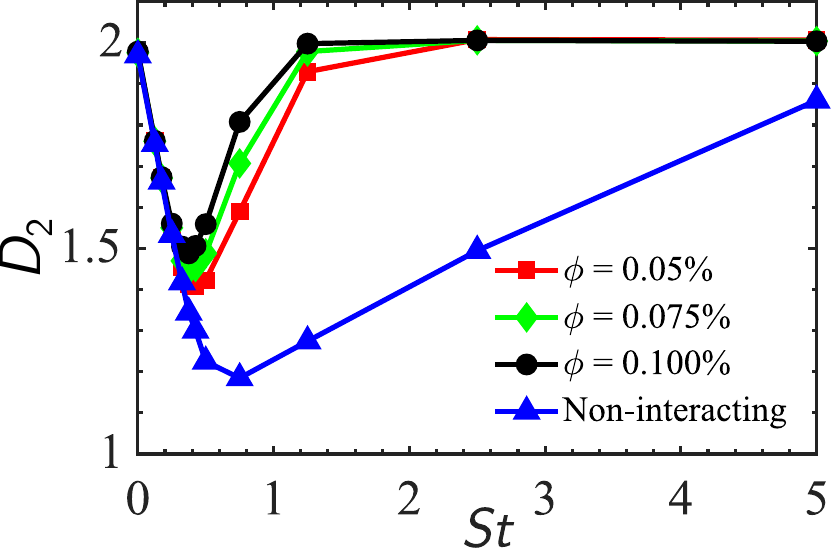}
  \caption{Correlation dimension $D_2$ \textit{vs} $St$ for non-interacting (blue triangles) and interacting particles 
    for ensembles for $\phi$, where $N_p$ varies (see legend). 
  }
  \label{D2}
\end{figure}

The effect of such inter-particle interactions is striking. In
Fig.~\ref{traj} we show representative snapshots of particle
positions, superimposed on the background vorticity field, for both
interacting (red) and non-interacting $V(r_{ij}) = 0$ (black)
particles.  In the absence of inertia ($St = 0$), unsurprisingly, the
difference between the two \textit{ensembles} is minute. However, for finite values of $St$ (panels (a)
and (b)), the particle distributions are strongly influenced by their
interactions. In particular, the nature of small-scale clustering is
significantly altered and, especially for $St > 1$ (panel~(b)),
interacting particles appear to be more homogeneously distributed than
the ones which are non-interacting.

\begin{figure*}
  \includegraphics[width=1.0\textwidth]{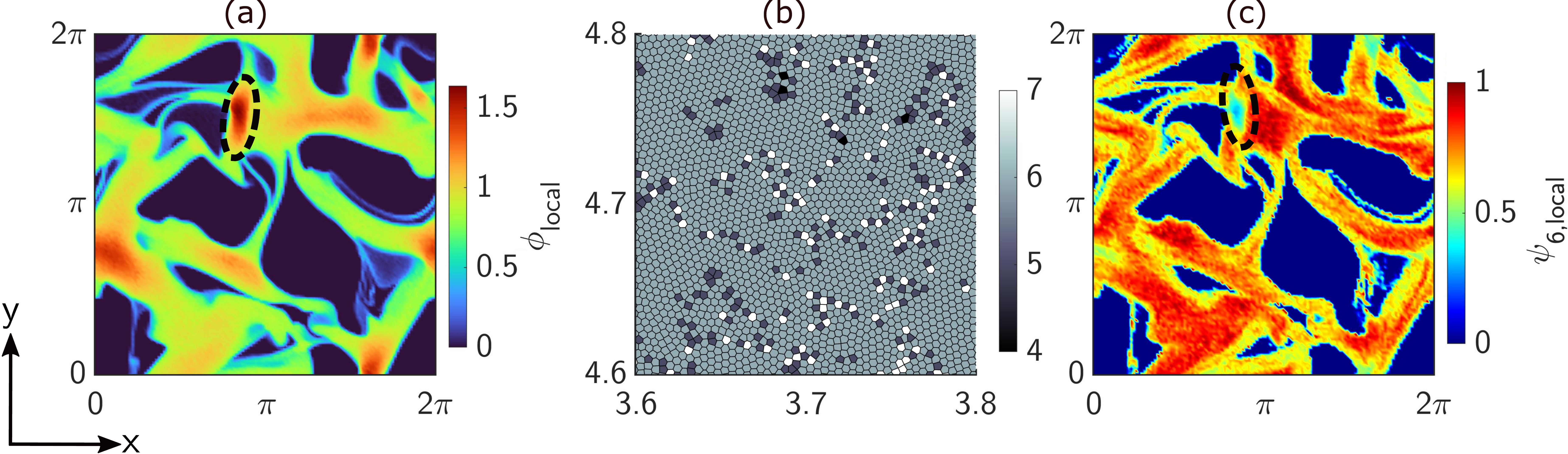}
  \caption{$\phi=50\%$. (a) Local packing fraction field
    $\phi_{\rm local}$ for the suspended grains, with $St \sim 1$.The hyper packed region is highlighted by dashed black oval shape.
    (b) The Voronoi tessellation along with the
    coordination number $z$ of each particle, shown as greyscale for a
    zoomed-in region having large $\psi_{6,\rm local}$, points to
    strong evidence of crystalline order with a hexagonal packing. Color-bar shows number of sides of each Voronoi cell. (c) Corresponding map to panel (a) for the local bond order parameter
	$\psi_{6,\rm local}$ with re-entrant melting region highlighted by dashed black line. (See Refs.~\cite{movie1-SI,movie1-youtube} for an 
	evolution of $\phi_{\rm local}$ and $\psi_{6,\rm local}$.)}
  \label{vorn_mono}
\end{figure*}

Inhomogeneities in the particle distribution are conveniently
characterized by the correlation dimension $D_{2}$, defined through
the probability of having two particles within a distance $r$, namely
$P^<_{2}(r) \sim r^{D_2}$, or equivalently through the small-scale
behavior of the radial distribution function $g(r) \sim r^{D_2-2}$.
Figure~\ref{D2} shows $D_2$ as a function of $St$ (including in
the non-interacting case), for various values of the packing fraction
$\phi = {N_p\sigma^2}/({16\pi})$. We fix $\sigma=5.0\times 10^{-4}$,
and choose $N_p=1\times 10^5$, $1.5\times 10^5$ and $2.0\times 10^5$,
to obtain $\phi= 0.05\%, 0.075\%$ and $0.10\%$, respectively. For a
given packing fraction and $St \ll 1$, the value of $D_2$, within
error-bars, are indistinguishable from the non-interacting particles.
This is because at such small values of $St$, there is hardly any
small-scale clustering and therefore such dilute suspensions are only
weakly affected by the short-range particulate interactions. However,
as particles cluster, i.e.\ for $St = \mathcal{O}(1)$, soft granular
repulsions dominate and, unlike the non-interacting ensemble, the
interacting particles spread more\,---\,as a result of the competing
interactions of vortical-ejection due to inertia and the short-range
inter-grain repulsive energy cost\,---\,and with a larger value of
$D_2$.  This effect also leads to a slight shifting to the left of the
value of $St$ where $D_2$ attains its minimum.  This effect is of
course accentuated with increasing packing fractions. For Stokes
numbers a bit larger than 1, although the centrifugal
vortical-ejection weakens, particles still tend to cluster in
straining zones.  However, the strength of the short-range repulsive
forces ensure that interacting particles spread out more and sample
the flow homogeneously, overcoming the bias due to inertia
(Fig.~\ref{traj}(b)). We see evidence of this in our measurements,
which show that for $St > 1$ the correlation dimension $D_2$
asymptotes to the physical dimension 2 much faster for the interacting
than for the non-interacting case. Furthermore, the 
minimum value of $D_2$ also increases as a function of $\phi$. Our results 
for packing fractions up to $\phi = 50\%$ (see Fig.~S1 in Ref.~\cite{SI})
shows clearly that the minima in the correlation dimension becomes shallower and shallower; 
however it is a matter of speculation if the minima does not exist at even 
higher levels of packing.

Thus, as observed and discussed, inter-particle interactions in particle-laden turbulent flows clearly
have an effect on the degree and nature of preferential
concentration. But are such particle interactions strong enough to
overcome turbulent mixing and nucleate crystalline structure? There are recent 
experimental studies which suggest that such aggregates may develop in homogeneous turbulence~\cite{JFM}. To
answer this question\,---\,and provide compelling evidence\,---\,it is
essential to work with a much larger packing fraction and particle
diameter, as is common in studies of granular
systems~\cite{gran-pack-1, gran-pack-2}.  (We however ensure that the
diameters are still much smaller than the fluid characteristic length
scale $l_\nu$ for our model to be valid.) We therefore choose
$\sigma = 5 \times 10^{-3}$, and different particle numbers, namely,
$N_p = 2\times 10^5$ ($\phi = 10\%$) and $N_p = 10^6$ ($\phi = 50\%$).

A useful indicator of how densely such particles are packed, due to
inertia and interactions, is to look at the local packing fraction
$\phi_{\rm local} = {N_\Delta\pi \sigma^2}/({4 \Delta^2})$, where
$N_\Delta$ is the number of particles in a small square of side
$\Delta = 4\,\delta x$, where $\delta x$ is the width of our Eulerian
grid.  In Fig.~\ref{vorn_mono}(a) we show a representative
pseudo-color snapshot of the local packing fraction for particles with
$St = 1$ ($\phi =
50\%$). As we would expect, in a given snapshot, there are regions which are
extremely dense and the local packing fraction far exceeds its average
$\phi$, with some regions hyper-packed due to the softness of the
particles.  We now examine the structure of these densely packed
regions by using the standard approach of \textit{Voronoi
  tessellation}. In Fig.~\ref{vorn_mono}(b) we show the Voronoi
construction, corresponding to a zoomed in region of panel (a);
furthermore we color, on a greyscale, each cell by the coordination
number $z$ (values shown in the adjacent colorbar) of the particle. We
find, surprisingly, that these soft particles do form hexagonal
lattices, as suggested by the predominance of $z=6$ (light grey) and
the cells of our Voronoi tessellation, with almost perfect crystalline
order.

To quantify this degree of crystallinity, we use the standard measure of the
bond order parameter $\psi_6$ ~\cite{psi61} for a given particle, 
$\psi_6(\textbf{r}_i)=\frac{1}{N_b}\left |\sum_{m=1}^{N_b}\exp{6\,\iota\, \theta_{mi}}\right |$
here $N_b$ is the number of nearest neighbors of the $i$th particle,
and $\theta_{mi}$ is the angle between the $x$-axis and the bond
joining the $i$th particle with the $m$th particle. Before we turn to
the full statistics of the bond order parameter, it is useful to first
look at a coarse-grained measure of this quantity. In analogy to our
definition of a local packing fraction, we define a local bond order
parameter
$\psi_{6,\rm
  local}=({1}/{N_\Delta})\sum_{i=1}^{N_\Delta}\psi_6(\textbf{r}_i)$.
The map for $\psi_{6,\rm local}$, corresponding and strongly correlated to the packing shown in
Fig.~\ref{vorn_mono}(a), is displayed in Fig.~\ref{vorn_mono}(c).
Such a coarse-grained description shows macroscopically large regions
with a very high value of $\psi_{6,\rm local}$ consistent with the
visual suggestion of crystallinity in Fig.~\ref{vorn_mono}(b). Also,
note that wherever the soft particles are hyper-packed, the
crystallinity is lost, and we have a {\em re-entrant melting} scenario
\cite{xu2016} highlighted by the dashed black oval in Figs.~\ref{vorn_mono}(a) and (c).

Starting from an initial spatially homogeneous particle homogeneous, 
we have observed that the inhomogeneous structures, as illustrated in
Fig.~\ref{vorn_mono}, emerge over a timescale $t \approx 5\tau_f$ after the turbulent
flow is switched on; please refer to Refs.~\cite{movie1-SI,movie1-youtube} for 
an animation of this evolution as 
well as what happens subsequently when the flow is switched off. In the
steady state, the crystalline aggregates are seen to be continuously
advected by the underlying flow, and therefore undergo 
breaking and coalescing at a constant rate as the straining zones 
evolve with time. Thus, in most cases, a typical crystalline aggregate
is transient in nature. To estimate the time scale on which a particular crystalline aggregate persists, 
we track a subset of particles from our simulations which form a particular crystalline structure. This subset of particles form 
a crystalline structure at $t=47\tau_f$, with most of the constituents having  $\psi_6 \sim 1$, and persists 
for time-scales of the order of a few $\tau_f$ before \textit{breaking away} to form 
new crystalline structures~\cite{movie2-SI,movie2-youtube}.

Further, we have also investigated what happens
when in steady state conditions, the underlying turbulent flow is suddenly switched
off abruptly. The suspended particles should also come to rest,  and it indeed does~\cite{movie1-SI,movie1-youtube},
over a timescale determined by $\tau_p$. The interesting finding is that the crystalline aggregation 
in the form of the spatially inhomogeneous structures remain intact; see Figs.~S4 (a) and (b) in the 
Supplementary Material~\cite{SI} as well as Refs.\cite{movie1-SI,movie1-youtube}.

\begin{figure}
  \includegraphics[width=1.0\columnwidth]{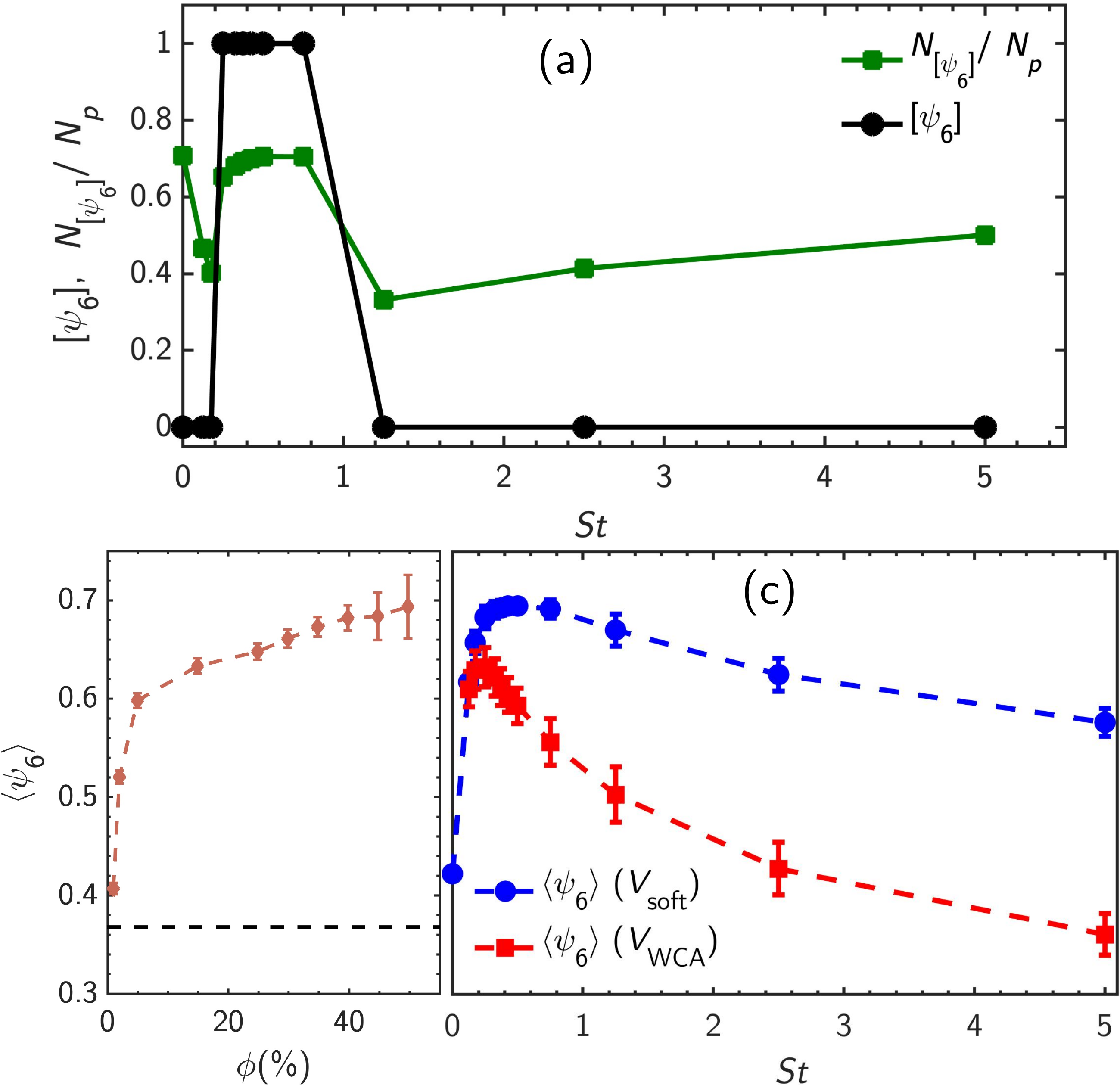}
  \centering
  \caption{(a) $\phi=10\%$. Variation of $[\psi_6]$ (black circles), the mode value of $\psi_6$ computed over all particles in configurations sampled in steady-state, and their associated
    fraction of particles $N_{[\psi_6]}/N_p,$ (green squares), with Stokes number   $St$. 
(b) For $St=1$, variation of mean value of $\psi_6$, $\langle\psi_6\rangle$ with $\phi$ showing a clear onset of crystalline aggregation, from the dilute random packing (horizontal dashed line).
(c) Comparison of $\langle\psi_6\rangle$ \textit{vs} $St$ for particles interacting via soft harmonic potential, shown with blue circles (for $\phi=10\%$,
    $N_p=2\times 10^5$ and $\sigma=5\times 10^{-3}$) and WCA potential, shown with red squares (for $\phi=25\%$,
	$N_p=5\times 10^5$ and $\sigma=4.4\times 10^{-3}$).  } 
\label{psi6_stat}
\end{figure}

The snapshots of Fig.~\ref{vorn_mono} naturally lead us to examine the
behaviour of $\psi_6$ as a function of $St$ for a reasonably high packing
fraction ($\phi\sim 10\%$). Given the non-equilibrium, spatio-temporal
variation of the advecting turbulent flow, as discussed above,  it is of course
natural that all the particles would not arrange themselves in a hexagonal
lattice. We therefore look at the \textit{mode} $[\psi_6]$, the value of
$\psi_6$ that occurs most frequently amongst all particles and over time. 
In the non-crystalline 
phase, the largest number of particles have a value close to 0 whereas in the crystalline 
phase this value tends to 1. 

In Fig.~\ref{psi6_stat}(a), we plot $[\psi_6]$ (black filled circles) as a
function of $St$ and see a remarkable behaviour.  For extremely small or large
values of the Stokes number, $[\psi_6] = 0$ whereas for values of $St$ around
1, where, as seen in Fig.~\ref{D2}, particles show significant preferential
concentration, $[\psi_6] = 1$. This behaviour is remarkable as it shows a sharp
transition between a crystalline and non-crystalline phase (as a function of
the Stokes number).  Of course such a characterisation makes sense only if
there is a macroscopically large fraction of the total particles which show
$[\psi_6]$ as reported in Fig.~\ref{psi6_stat}(a). We thus calculate the
fraction of particles $N_{[\psi_6]}/N_p$ (along with their errorbars calculated
over time) having $\psi_6=[\psi_6]$ (green square symbols).
At a practical level, $\psi_6$ of course varies from 0 to 1; therefore 
it is convenient for us to bin values of $\psi_6$. In particular, we count all particles 
with values of $0.0\le \psi_6 \le 0.2$ as $\psi_6 = 0$ and those in the range 
 $0.8\le \psi_6 \le 1.0$ as $\psi_6 = 1.0$. We have checked that slight variations in this range does 
 not change our results (Fig.~\ref{psi6_stat}).
Not surprisingly, in the homogeneously dispersed phase ($St =
0$) we find close to $70\%$ of the particles are associated with $\psi_6 =
0.0$. However as $0 < St \ll 1.0$, the spheres tend to weakly cluster leading
to a greater variation in $\psi_6$. Thus for this range of Stokes numbers,
while $[\psi_6] = 0$, there is a decrease in $N_{[\psi_6]}/N_p$ as seen in
Fig.~\ref{psi6_stat}(a). It is only when the Stokes numbers reach values where
particles preferentially concentrate most strongly (see Fig.~\ref{D2} for
comparison), that $[\psi_6] = 1.0$ and a consequent rise and plateauing of the
value of $N_{[\psi_6]}/N_p$. Thus, in this cryastalline phase where
$[\psi_6]=1$, we immediately see (Fig.~\ref{psi6_stat}(a)) that a majority of
particles arrange themselves in perfect hexagonal order, indicating that indeed
there is a dynamical structural transition with Stokes number. For higher Stokes numbers, the particles start to disperse
more homogeneously, and the crystalline phase is lost, for reasons discussed
before, leading to $[\psi_6]=0$.  While for finite Stokes numbers, there are
still great variations in the overall values of $\psi_6$ associated with each
particle, as the Stokes number becomes larger and larger, the majority fraction
$N_{[\psi_6]}/N_p$ of those with $\psi_6 = 0$ increases. The results from our
simulations, shown in Fig.~\ref{psi6_stat}(a), are consistent with this
theoretical understanding of such soft-sphere suspensions.

In the crystalline phase, 
the variation of the degree of crystallinity $\langle \psi_6 \rangle$ with the packing fraction 
is shown in Fig.~\ref{psi6_stat}(b). Clearly, and as conjectured, $\langle \psi_6 \rangle$ increases with 
$\phi$; remarkably even for very dilute suspension, we do see an average crystalline order with $\langle \psi_6 \rangle$ 
significantly higher than the value for random packing.

While the two curves in Fig.~\ref{psi6_stat}(a) give
quantitative evidence for the emergence of a macroscopic crystalline order in
an ensemble of particles in a turbulent flow, it is useful to also examine 
the mean bond order parameter $\langle \psi_6 \rangle$ as a function
of the Stokes number. The non-monotonic behaviour of this measure, seen in the curve for $V_{\rm soft}$ in Fig.~\ref{psi6_stat}(c),  is consistent
with that seen for $[\psi_6]$.  Of course the average value does not switch
between 0 and 1 because of local fluctuations in the particle arrangement. For
the range of Stokes numbers where crystalline structures are observed, we find
good evidence for onset of crystallinity by measuring $\langle \psi_6 \rangle$ as a function of $\phi$ 
(Fig.~\ref{psi6_stat}(c)).

Since the results reported so far have been obtained for soft particles,
whereby two particles can overlap/penetrate to mimic deformation 
at contact for real grains or droplets, it is
important to check if this phenomenon is \textit{universal} and, in particular,
persists even for hard-core particles, where the repulsive forces
will be stronger at contact and thereby affect local packing. In order to do so, we performed simulations
where we model the \textit{hard cores} particles via the 
Weeks-Chandler-Anderson (WCA) inter-particle potential~\cite{wca} \--\ $V_{\rm WCA}(r_{ij})= 4
\epsilon[(\sigma/r_{ij})^{12}-(\sigma/r_{ij})^{6}]+\epsilon$ for
$r_{ij}\leq2^{\frac{1}{6}}{\sigma}$ and $0$ otherwise. For this system, we again
measure $\langle \psi_6 \rangle$ (in Fig.~\ref{psi6_stat} (b), red
squares), with a packing fraction twice as large as that used for soft disks,
as a function of $St$. Also see Fig.~S2 in Supplementary Information \cite{SI} for
local maps for the WCA particles, similar to what has been shown 
in Fig.~\ref{vorn_mono} for the harmonic disks. Our measurements show that the degree of maximum
crystallinity, at $St \sim 1$, is marginally smaller than that for particles
with soft interactions and that $\langle \psi_6 \rangle$ falls off faster
as a function of $St$ for the assembly of \textit{hard} disks as compared to \textit{soft}
disks. Nevertheless, the basic mechanism of the formation of crystalline aggregates 
seem to be unchanged even for the extreme case of the WCA potential. 

Another consideration is the polydispersity of the granular
particles and it's effects on the crystalline aggregates~\cite{John2017,
Pablo2019, Pablo2020}. To study this we use a bi-disperse assembly of
particles with the particle radius ratio of 1.4~\cite{o2003jamming}. As expected, for such 
mixtures, the spatial extent of local crystallinity diminishes significantly when compared 
to the mono-component system (see Fig.~S3 of the Supplementary Material~\cite{SI}). 
Nevertheless, even for such bi-disperse suspension, dense  packings leading to locally jammed
structures, induced by the carrier 
flows, do still occur. Indeed this issue along with the 
role played by the energy scale $\epsilon$ needs a detailed study in future.

In conclusion, we have shown that two-dimensional particle-laden
turbulent flow, lead to crystalline self-assembly, albeit spatially inhomogeneous, because of the
complementary effects of drag-induced preferential concentration and
inter-particle interactions \cite{foot1}. The central role played by
particle inertia is apparent in the sharp transition between
crystalline aggregates as a function of the Stokes
number. Our work also shows that elasticity of the
particles lead to a modification in the nature of preferential
concentration of heavy inertial particles which lie at the heart of
several natural and industrial processes.

We also demonstrate that these crystalline aggregates survive even after the
turbulent flow is switched off (Fig.~S4 of the Supplementary Material~\cite{SI} and Refs.~\cite{movie1-SI,movie1-youtube}) 
by setting $\textbf{u}(\textbf{x}_i,t)=0$ in
Eq.~\ref{stokes}. Thus, this provides a route to engineer such inhomogeneous
particulate assemblies, for example via drying of the suspending fluid, as it
often does in nature. We further note that the chosen value
of $\epsilon$ sets the energy scale such that potential energy is comparable to
the kinetic energy as shown in Fig.~S3 of the Supplementary Material~\cite{SI}.

It is also important to keep in mind that, given that there have
hardly been any studies of particulate structures in a turbulent flow,
our work focuses on the most simple and general framework as commonly
used in turbulent transport problems. The most important
simplification that we have used is to ignore the feedback of the
particles on the flow, as well as lubrication forces or the effect of
porosity in the packed structures. It has been shown in an earlier
work ~\cite{Saw-PoF} that a one-way coupled model for Stokesian
particles is a valid assumption in turbulent flows. Furthermore, it
was shown in~\cite{sticky}, while studying the problem of elastic
collisions in particle-laden turbulent flows, that at least in the
small Stokes limit, the effect of short-range lubrication was merely
the renormalisation of the effective relaxation time. However, it
should be left for future work to actually examine in detail the role
of lubrication and porosity in stabilizing such crystalline
structures.

Finally, we note that such inhomogeneous crystalline aggregation has recently
been observed in assemblies of self-propelled particles \cite{palacci}.
However, the morphology of the clusters in the case 
of the turbulence led aggregation is perhaps very different,
and this needs to be explored further. However, in a broader
context, the diverse non-equilibrium routes for such self-assembly
opens up fascinating avenues. 

\begin{acknowledgments}

We thank J. R. Picardo and S. Puri for useful suggestions and discussions.
JB and SSR acknowledge the support of the Indo-French Centre for Applied
Mathematics (IFCAM). 
 SSR acknowledges the support of SERB-DST (India) projects MTR/2019/001553,
	STR/2021/000023, CRG/2021/002766 and the DAE, Govt. of India, under
	project no.  12-R\&D-TFR-5.10-1100 and project no.  RTI4001
	PC acknowledges CEFIPRA Grant No. 5604-1 for financial
support.  This research was supported in part by the International Centre for
Theoretical Sciences (ICTS) during a visit of PC for participating in the program -
Entropy, Information and Order in Soft Matter (Code: ICTS/eiosm2018/08).  The
simulations were performed on the ICTS clusters {\it Mowgli} {\it Mario}, and
{\it Tetris}, as well as the work stations from the project ECR/2015/000361:
{\it Goopy} and {\it Bagha}.

\end{acknowledgments}

\bibliography{ref_soft_pt}

\end{document}


\title{Supplemental Information for ``Turbulent Route to Two-Dimensional Soft Crystals"}
\author{Mohit Gupta}
\email{mohit.gupta9607@gmail.com}
\affiliation{\ictsaddress}
\affiliation{School of Physics and Astronomy, University of Minnesota, Minneapolis, Minnesota 55455, USA}

\author{Pinaki Chaudhuri}
\email{pinakic@imsc.res.in}
\affiliation{\imscaddress}
\author{J\'er\'emie Bec}
\email{jeremie.bec@mines-paristech.fr}
\affiliation{\inria}
\affiliation{\minesaddress}
\author{Samriddhi Sankar Ray}
\email{samriddhisankarray@gmail.com}
\affiliation{\ictsaddress}

\maketitle

\section{Supplementary Figures }

\subsection{$D_2$ at higher packing fractions}
\begin{figure*}[!htb]
\includegraphics[width=0.8\textwidth]{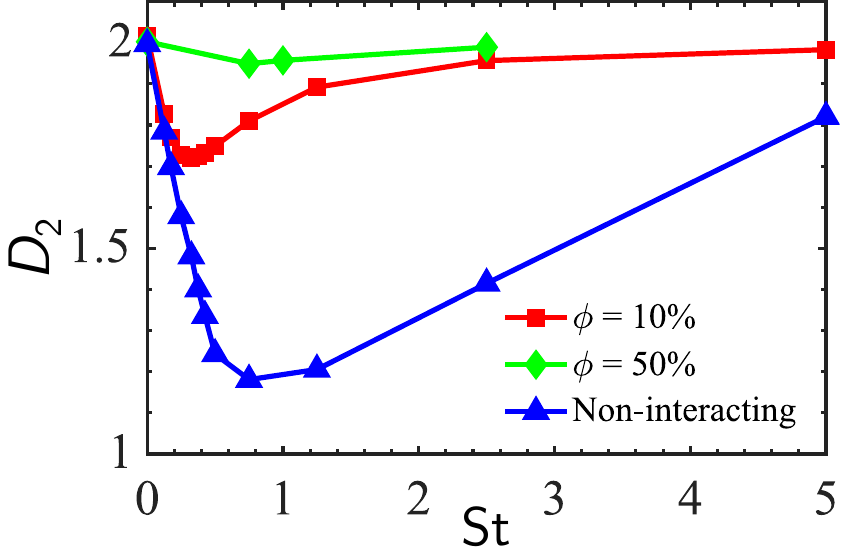}
\caption{{\em Soft Harmonic disks.} Correlation dimension $D_2$ \textit{vs} ${\rm St}$ for non-interacting (blue triangles) and interacting particles, where $N_p$ is fixed
and $\phi$ is varied by changing the particle diameter. (The error-bars are smaller than the symbols size).}
\label{D2_phi50}
\end{figure*}

The minimum value of obtained $D_2$ as a function of ${\rm St}$ increases as packing fraction increases. To study weather the minim in the $D_2$ disappears at a high enough packing fraction we show data obtained at higher packing fractions in Fig.~\ref{D2_phi50}.

\subsection{Hard-core particles: Weeks-Chandler-Anderson Potential}

\begin{figure*}[h]
\includegraphics[width=1\textwidth]{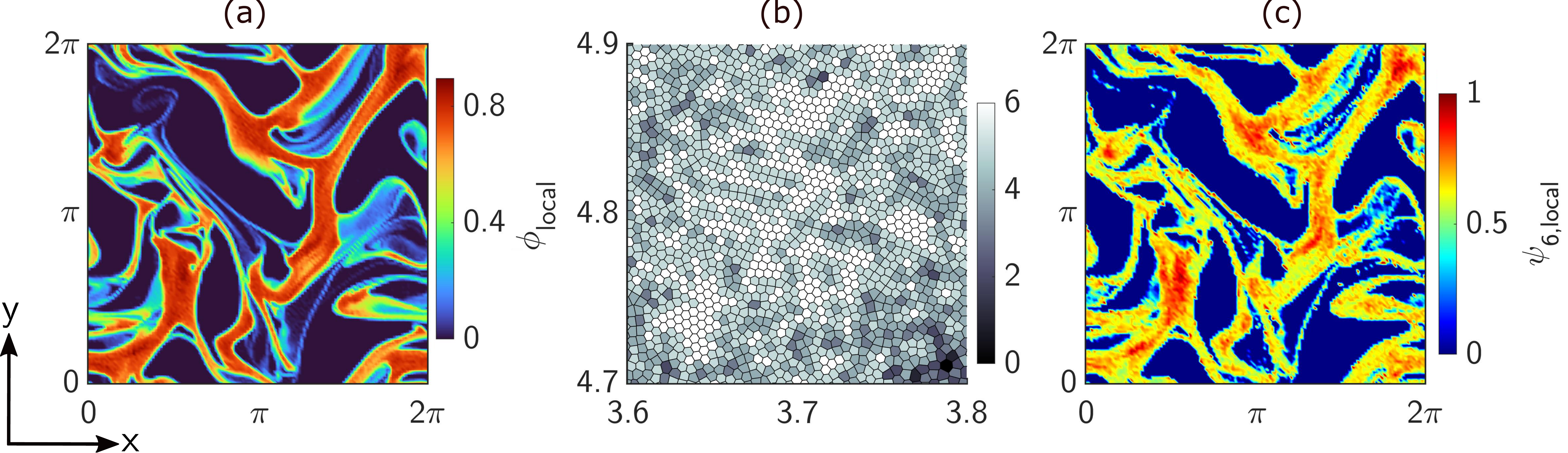}
\caption{$\phi=25\%$. Particles interacting via WCA potential. (a) Local packing fraction field
    $\phi_{\rm local}$ for the suspended grains, with ${\rm St} \sim 1$. (b) The Voronoi tessellation along with the
    coordination number $z$ of each particle, shown as greyscale for a
    zoomed-in region having large $\psi_{6,\rm local}$ indicates
    a weak crystalline order compared to the soft harmonic potential. Color-bar shows number of sides of each Voronoi cell. (c) Corresponding map for the local bond order parameter
    $\psi_{6,\rm local}$.}
\label{s1}
\end{figure*}

To study the response of an assembly of particles, with hard-core like interactions, we consider a 
monodisperse system where particles interact via the Weeks-Chandler-Anderson (WCA) potential (see main text for the functional form). 
We study the structures at $\phi=25\%$, using  $N_p=5.0\times10^5$ and $\sigma=0.0044$ with effective diameter being 
$2^{1/6}\times\sigma=0.005$. As illustrated in Fig.\ref{s1}, we observe that the degree of crystalisation is less compared to the case of \textit{harmonic disks}.

\subsection{Bi-disperse assembly of soft particles}
\begin{figure*}
\includegraphics[width=1\textwidth]{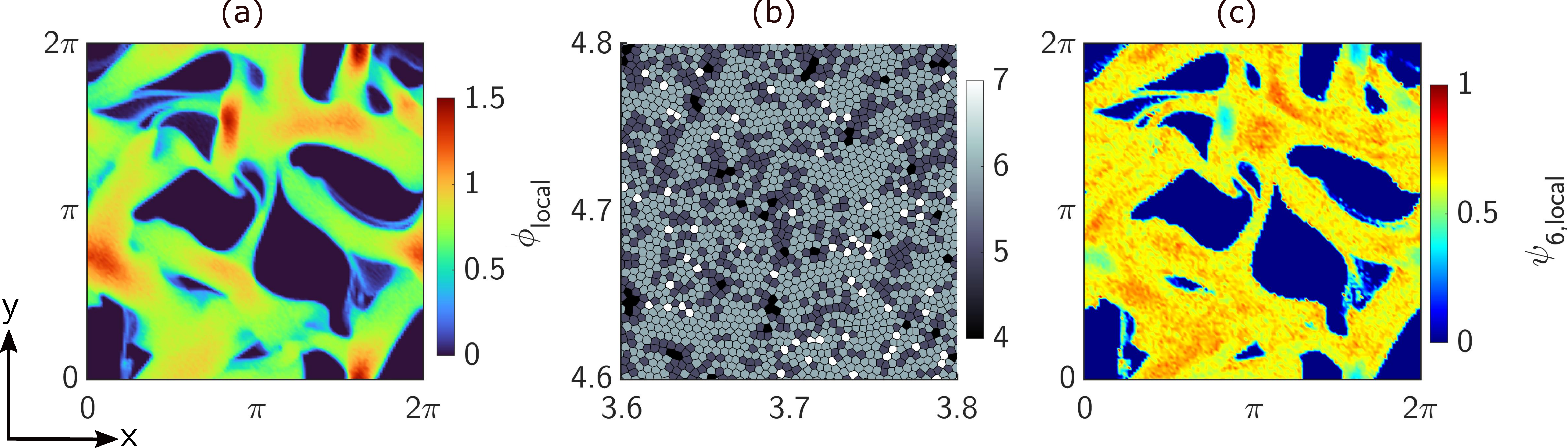}
\caption{$\phi=60\%$. Bi-disperse assembly of particles interacting via soft potential. (a) Local packing fraction field
    $\phi_{\rm local}$ for the suspended grains, with ${\rm St} \sim 1$. (b) The Voronoi tessellation along with the
    coordination number $z$ of each particle, shown as greyscale for a
    zoomed-in region having large $\psi_{6,\rm local}$. Color-bar shows number of sides of each Voronoi cell.
    (c) Corresponding map for the local bond order parameter
    $\psi_{6,\rm local}$ indicates
    a weak crystalline order compared to the mono-disperse solution.}
\label{s2}
\end{figure*}

To study the effect of size dispersity on the structure formation, we perform simulations of a
bi-disperse (50:50) system, with a size ratio of $1.4$. Here particles interact via the soft harmonic potential. We study the structures using  $N_p=10.0\times10^5$,
with the two sizes being $\sigma_1=0.005$ and $\sigma_2=0.007$, giving $\phi=60\%$. As illustrated in Fig.\ref{s2}, we observe that the spatial extent 
of crystalisation is significantly less compared to the monodisperse case. To isolate the effect of soft-particle interactions we set the Stokes number to be the same for both radius, which is equivalent to a change in the density of the particles.

\subsection{Particle assembly with $\textbf{u}=0$}
We have looked at the particle assembly after we \textit{turn off} the flow, i.e we have set $\textbf{u}=0$ after the system has reached steady state. As shown in the supplementary movie and seen from Fig.~\ref{s3} (c), the system freezes as mean kinetic energy $\langle E_k\rangle\sim 0$ and mean potential energy $\langle E_p\rangle\sim 0$ in time $t>\tau_p$. As particles come to a halt the order in the system remains stable (Fig.~\ref{s3} (b)).  
\begin{figure*}
\includegraphics[width=1\textwidth]{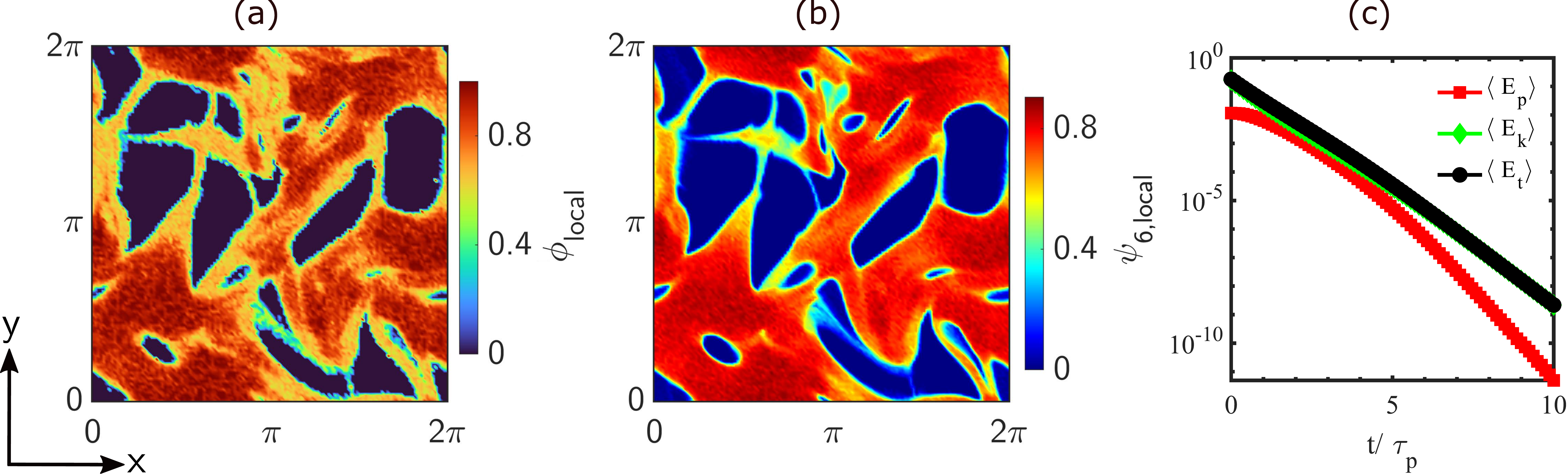}
\caption{$\phi=50\%$. Particle assembly after the flow is abruptly stopped. (a) Local packing fraction field
    $\phi_{\rm local}$ for the suspended grains, with ${\rm St} \sim 1$.
    (b) Corresponding map for the local bond order parameter
    $\psi_{6,\rm local}$. (c) Mean kinetic energy $\langle E_k\rangle$ (green circles), mean potential energy$\langle E_p\rangle$ (red circles) and total energy $\langle E_t\rangle$ (blue circles) of particles as function of time after the flow is stopped.}
\label{s3}
\end{figure*}

\section{Supplementary Movie}
\subsection{Particle Evolution}
We show time evolution of the system of  $N_p=10^6$ particles, having diameter $\sigma=5\times10^{-3}$ which gives $\phi=50\%$, interacting via harmonic interactions, with ${\rm St}=1$. We start from a spatially homogeneous distribution of particles.  In the movie, the left panel shows the local packing fraction field $\phi_{\rm local}$ for the suspended grains, and the right panel shows the corresponding map for the local bond order parameter $\psi_{6,\rm local}$. As the system evolves, for $t > \tau_f$, we obtain preferentially concentrated state where particles are concentrated in the straining zones, causing large density fluctuations. This leads to local crystalline order in the system as we can see from the value of $\psi_{6,\rm local}$.  

We \textit{turn off} the flow at $t=75\tau_f$ and it can be seen that the system \textit{freezes} and the particles obtain there equilibrium positions in $t>\tau_p$. The video is also available at \cite{movie_1}. 

\subsection{Cluster time scale}
We show time evolution of a subset of particles from our simulations ($N_p=10^6$, $\sigma=5\times10^{-3}$). The particles are colored according to their instantaneous $\psi_6$, with ${\rm St}=1$. As these particles move they form a crystalline structure at $t=47\tau_f$. After a few $\tau_f$ this structure breaks up and constituent particles evolve in time to form different clusters. The video is also available at \cite{movie_2}.